\documentclass[12pt,epsf]{article}
\usepackage{amsmath,amssymb,graphicx,epsfig,latexsym}
\numberwithin{equation}{section}

\newcommand{\be}{\begin{equation}}
\newcommand{\ee}{\end{equation}}

\newcommand{\beqa}{\begin{eqnarray}}
\newcommand{\eeqa}{\end{eqnarray}}
\newcommand{\nn}{\nonumber}

\def\boxit#1{\vbox{\hrule\hbox{\vrule\kern8pt
\vbox{\hbox{\kern8pt}\hbox{\vbox{#1}}\hbox{\kern8pt}}
\kern8pt\vrule}\hrule}}
\def\mathboxit#1{\vbox{\hrule\hbox{\vrule\kern8pt\vbox{\kern8pt
\hbox{$\displaystyle #1$}\kern8pt}\kern8pt\vrule}\hrule}}

\def\IB{\relax\hbox{$\inbar\kern-.3em{\rm B}$}}
\def\IC{\relax\hbox{$\inbar\kern-.3em{\rm C}$}}
\def\ID{\relax\hbox{$\inbar\kern-.3em{\rm D}$}}
\def\IE{\relax\hbox{$\inbar\kern-.3em{\rm E}$}}
\def\IF{\relax\hbox{$\inbar\kern-.3em{\rm F}$}}
\def\IG{\relax\hbox{$\inbar\kern-.3em{\rm G}$}}
\def\IGa{\relax\hbox{${\rm I}\kern-.18em\Gamma$}}
\def\IH{\relax{\rm I\kern-.18em H}}
\def\IK{\relax{\rm I\kern-.18em K}}
\def\IL{\relax{\rm I\kern-.18em L}}
\def\IP{\relax{\rm I\kern-.18em P}}
\def\IR{\relax{\rm I\kern-.18em R}}
\def\IZ{\relax\ifmmode\mathchoice
{\hbox{\cmss Z\kern-.4em Z}}{\hbox{\cmss Z\kern-.4em Z}}
{\lower.9pt\hbox{\cmsss Z\kern-.4em Z}} {\lower1.2pt\hbox{\cmsss
Z\kern-.4em Z}}\else{\cmss Z\kern-.4em Z}\fi}

\def\II{\relax{\rm I\kern-.18em I}}

\def\CG {{\cal G}}

\def\CO {{\cal O}}

\def\Lie{{\rm Lie}}

\begin{document}

\setlength{\baselineskip}{7mm}
\begin{titlepage}
\title {Closed Algebras for Higher Rank, non-Abelian Tensor Gauge Fields}
\maketitle
\begin{center}
\author{Spyros Konitopoulos\footnote{spykoni@gmail.com} }\\
\begin{abstract}
A systematic method is presented for the construction and classification of 
algebras of gauge transformations for arbitrary high rank tensor gauge fields. For every tensor gauge 
field of a given rank, the gauge transformation
will be stated, in a generic way, via an {\it ansatz} that contains all the possible terms, with arbitrary
coefficients and the maximum number of tensor gauge functions. 
The requirement for the closure of the algebra will prove to be restrictive, but, nevertheless,
leave a variety of choices. Properly adjusting the values of the initial coefficients and imposing restrictions
on the gauge functions, one can, one the one hand, recover all the, so far, analysed algebras and on
the other, construct new ones. The presentation of a brand new algebra for tensor gauge
transformations is the central result of this article. 
\end{abstract}
\end{center}
\end{titlepage}
\newpage
\pagestyle{plain}

\section {Introduction}
The investigation on theories describing fields of arbitrary high spin dates back to 1936, when {\it P. Dirac} formulated the Relativistic Wave Equations for massless and massive fermions of spin greater that 1/2  \cite{Dirac: 1936}. His work was extended by {\it M. Fierz} and {\it W. Pauli}, who followed a field theoretical approach to include bosons of higher spin \cite{Fierz-Pauli: 1939}, and tuned a few months later \cite{De Wet: 1940}. All the attempts gave elegant results for the free particle case, but did not succeed to include the electromagnetic interactions in a satisfactory and free of ambiguities unified framework. So far, the physical requirements for constructing higher spin theories was the invariance under {\it Poincar\'e} transformations and the positivity of the energy after quantisation \cite{Fronsdal: 1978}. The revolutionary, group theoretical approach carried out in \cite{Wigner: 1939, Bargmann-Wigner: 1948}, together with the establishment of the {\it Principle of Invariance under Gauge Transformations} \cite{Yang-Mills: 1954}, opened the way for the introduction of new models for the description of higher spin fields with less restrictive principal requirements 
\cite{Fronsdal: 1957, Fronsdal: 1978, Fang-Fronsdal: 1978}. 
Nevertheless, a satisfactory and self contained, interacting higher spin field theory was yet to be formulated both for the abelian (electromagnetic) and non-abelian interactions (weak, strong). 

On 2005 it was proposed that the description of higher spin particles can be carried out 
following a natural extension of the {\it Yang-Mills} ({\it Y-M}) principle to include higher rank 
bosonic and fermionic fields \cite{Savvidy: 2005}. The theory was developed quite extensively under the name
Tensor Gauge Field Theory, has been theoretically tested with regards to self consistence issues \cite{Savvidy: 2006b, Spyros-Savvidy: 2008}, properly analysed and extrapolated to provide both phenomenological  and theoretical predictions \cite{Savvidy: 2006c, Spyros-Fazio-Savvidy: 2009, Spyros-Savvidy: 2009, Savvidy: 2010, Spyros-Savvidy: 2014, Savvidy: 2014, Spyros-Savvidy: 2016}
and served as a basis for further developments \cite{Salgados: 2017}. 

As far as one departs from the {\it Y-M} case, where the mediators of the forces are described by
vector gauge fields, one faces the challenging problem of defining new algebras for tensor gauge transformations.
Under the light and guidance of the {\it Y-M} principle, the starting point for the formulation of a consistent 
higher spin theory is the extension of gauge transformations to higher rank tensor fields. 
Fortunately, this investigation is facilitated by the requirement that new proposed transformations
are legitimate candidates as long as they form a closed algebraic structure and hence belong to 
the class of {\it Lie} algebras. Quantitatively this is translated to the imperative that the 
commutator of two infinitesimal successive transformations belongs to the same set of
transformations.

In the following section, we give a brief review of all closed algebras that have been proposed under the
framework of Tensor Gauge Field Theory and make some remarks concerning their internal structure. 
In section 3, we present a general method that will, one the one hand, allow the embedding of the
previously mentioned algebras in a generalised framework and on the other, facilitate the investigation
of new ones. Finally, we conclude with a review of the new algebra constructed by the introduced method and 
comment on the potentials of its usage.

\section {Known Closed Algebras}
In recent papers, \cite{Savvidy: 2005, Savvidy: 2006, Savvidy-Barret: 2007, Savvidy-Guttenberg: 2008, Savvidy-Georgiou: 2010}, a number of gauge transformations for bosonic tensor fields of arbitrary high rank
have been constructed and proven to form a closed algebraic structure. 
Each of these algebras was constructed to serve a particular purpose. 
The first that appeared historically, under the framework of Tensor Gauge Field Theory, served as the building block
for the construction of field strength tensors, by the aid of which gauge invariant {\it Langrangians} for higher
rank fields were presented. We will call these transformations the standard ones, to discriminate from the others
that followed. The dual gauge transformations, appeared in two versions, and elucidated the fact that two seemingly 
different algebras can be equivalent and thus related by a similarity transformation. 
The symmetric gauge Transformations was part of an attempt to formulate a theory solely for the irreducible 
representations of the {\it Poincar\'e} group which proved to be technically impossible. 
Finally, the algebra of the gauge transformations used for the construction of new topological invariants in
higher dimensions is our last example of extended algebras. We will call these transformations, conjugate,
for classification purposes. 

\subsection {Standard Extended Gauge Transformations}
The algebra of standard extended gauge Transformations appeared for the first time in 
\cite{Savvidy: 2005}, as the fundamental building block of non-Abelian Tensor Gauge Field Theory
\cite{Savvidy: 2006, Savvidy: 2006b}. It concerns higher rank bosonic fields, $A_{\mu\lambda_1\lambda_2\dots\lambda_s}$, 
which are by construction  symmetric under the permutation of their lambda indices,
but bear no symmetry with respect to the index $\mu$. 
Besides, the tensor gauge functions, $\xi_{\lambda_1\lambda_2\dots\lambda_s}$,
needed for the definition of the gauge transformations of $A_{\mu\lambda_1\lambda_2\dots\lambda_s}$,
are totally symmetric\footnote{The tensor gauge fields and the tensor gauge functions carry a colour index which
enumerates the independent generators of the underlying $SU(N)$ {\Lie} algebra. Hence, both the quantities
are summation shortcuts over the fundamental representation matrices of the generators of the {\it Lie} algebra,\\
$A_{\mu\lambda_1\lambda_2\dots\lambda_s}=A^a_{\mu\lambda_1\lambda_2\dots\lambda_s}L^a$ and 
$\xi_{\lambda_1\lambda_2\dots\lambda_s}=\xi^a_{\lambda_1\lambda_2\dots\lambda_s}L^a$,
where $[L^a, L^b]=if^{abc}L^c$.}.

For successively higher rank fields, the gauge transformations are given below,
\beqa\label{extended gauge transformations}
\delta_{\xi}A_{\mu}&=&\partial_{\mu}\xi-ig[A_{\mu}, \xi] \nn \\
\delta_{\xi}A_{\mu\lambda}&=&\partial_{\mu}\xi_{\lambda}-ig\left( [A_{\mu}, \xi_{\lambda}]+[A_{\mu\lambda}, \xi]\right) \nn \\
\delta_{\xi}A_{\mu\lambda_1\lambda_2}&=&
\partial_{\mu}\xi_{\lambda_1\lambda_2}-ig\left( [A_{\mu}, \xi_{\lambda_1\lambda_2}]+[A_{\mu\lambda_1}, \xi_{\lambda_2}]+
[A_{\mu\lambda_2}, \xi_{\lambda_1}]+[A_{\mu\lambda_1\lambda_2}, \xi]\right)\nn \\
&&\dots\dots  \nn \\
\delta_{\xi}A_{\mu\lambda_1\dots\lambda_s}
&=&\partial_{\mu}\xi_{\lambda_1\dots\lambda_s}-ig\sum_{i=0}^{s}\sum_{Ps}[A_{\mu\lambda_1\dots\lambda_i}, 
\xi_{\lambda_{i+1}\dots\lambda_s}].
\eeqa
The first is the well known {\it Y-M} gauge transformation for the vector field, while the rest define
the way the higher rank bosonic tensors transform. In particular, the non-homogenous part of the
transformations involves differentiation in terms of the first index  $\mu$.
As one can show, they define an infinite dimensional gauge group, $\CG$, with a closed algebraic structure \cite{Savvidy: 2006},
\beqa
[\delta_{\xi},\delta_{\eta}]A_{\mu\lambda_1\lambda_2\dots\lambda_s}=-ig\delta_{\zeta}
A_{\mu\lambda_1\lambda_2\dots\lambda_s}
\eeqa
where,
\beqa
\zeta&=&[\xi,\eta]\nn \\
\zeta_\lambda&=&[\xi,\eta_\lambda]+[\xi_\lambda,\eta]\nn \\
\zeta_{\lambda_1\lambda_2}&=&[\xi,\eta_{\lambda_1\lambda_2}]
+[\xi_{\lambda_1},\eta_{\lambda_2}]+[\xi_{\lambda_2},\eta_{\lambda_1}]+
[\xi_{\lambda_1\lambda_2}, \eta]\nn \\
&\dots\dots&\nn \\
\zeta_{\lambda_1\dots\lambda_s}&=&[\xi,\eta_{\lambda_1\dots\lambda_s}]+
\sum_{i=1}^s{[\xi_{\lambda_i},\eta_{\lambda_1\dots\lambda_{i-1}\lambda_{i+1}\dots\lambda_s}]}+\dots+
[\xi_{\lambda_1\dots\lambda_s},\eta]
\eeqa
The above algebra allowed for the definition of consistent field strength tenors which gauge transform homogeneously, 
for the construction of two classes of gauge invariant {\it Lagrangians} for each rank of tensor gauge fields, 
and for a fundamental extension of the {\it Poincar\'e} group \cite{Savvidy: 2010b}.

\subsection {Dual Extended Gauge Transformations}
Parallelly to the previous transformations, one can define complementary ones which ended up to be
equivalent with the former. 
They have been formulated in two different versions \cite{Savvidy-Barret: 2007, Savvidy-Guttenberg: 2008}, 
which we will shortly review.  
\subsubsection{1st Version}
In the first version \cite{Savvidy-Barret: 2007}, the roles of the $\mu$ and the $\lambda$ indices are 
properly interchanged so that the inhomogeneous terms entail derivatives with 
respect to each of the latter indices. Up to the tensor of third rank, the transformations are listed below,
\beqa\label{dual}
\tilde{\delta}_{\xi}A_{\mu}&=&\partial_{\mu}\xi-ig[A_{\mu}, \xi] \nn \\
\tilde{\delta}_{\xi}A_{\mu\lambda}&=&\partial_{\lambda}\xi_{\mu}-ig\left(
[A_{\lambda}, \xi_{\mu}]+[A_{\mu\lambda}, \xi]\right) \nn \\
\tilde{\delta}_{\xi}A_{\mu\lambda_1\lambda_2}&=&
\partial_{\lambda_1}\xi_{\mu\lambda_2}+\partial_{\lambda_2}\xi_{\mu\lambda_1}
-ig\big( 
[A_{\lambda_1}, \xi_{\mu\lambda_2}]+[A_{\lambda_2}, \xi_{\mu\lambda_1}]+
[A_{\mu\lambda_1}, \xi_{\lambda_2}]+[A_{\mu\lambda_2}, \xi_{\lambda_1}]+\nn \\
&&+[A_{\lambda_1\lambda_2},\xi_\mu]+[A_{\lambda_2\lambda_1},\xi_\mu]+
[A_{\mu\lambda_1\lambda_2},\xi]
\big)\nn \\
&&\dots\dots  \nn \\
\eeqa
It has been proved \cite{Savvidy-Barret: 2007}, that the above transformations 
form a closed algebraic structure. Indeed, the commutator of two successive transformations leads to a third one,
\beqa
[\tilde{\delta}_{\xi},\tilde{\delta}_{\eta}]A_{\mu\lambda_1\dots\lambda_s}=-ig\tilde{\delta}_{\zeta}
A_{\mu\lambda_1\dots\lambda_s},
\eeqa
with
\beqa
\zeta&=&[\xi,\eta]\nn \\
\zeta_{\lambda}&=&[\xi,\eta_\lambda]+[\xi_\lambda,\eta]\nn \\
\zeta_{\lambda_1\lambda_2}&=&[\xi,\eta_{\lambda_1\lambda_2}]+
[\xi_{\lambda_1},\eta_{\lambda_2}]+[\xi_{\lambda_2},\eta_{\lambda_1}]+
[\xi_{\lambda_1\lambda_2},\eta]\nn \\
&\dots\dots&\nn \\
\zeta_{\lambda_1\dots\lambda_s}&=&[\xi,\eta_{\lambda_1\dots\lambda_s}]+
\sum_{i=1}^s{[\xi_{\lambda_i},\eta_{\lambda_1\dots\lambda_{i-1}\lambda_{i+1}\dots\lambda_s}]}+\dots+
[\xi_{\lambda_1\dots\lambda_s},\eta]
\eeqa
Besides this, in the same article, it has been shown that the introduction of the transpose tensor gauge fields,
\beqa\label{transpose1}
\tilde{A}_{\mu\lambda}&=&A_{\lambda\mu}\nn \\
\tilde{A}_{\mu\lambda_1\lambda_2}&=&
{1\over 2}\left(A_{\lambda_1\mu\lambda_2}+A_{\lambda_2\mu\lambda_1}\right)
-{1\over 2}A_{\mu\lambda_1\lambda_2}\nn \\
&\dots\dots&\nn \\
\tilde{A}_{\mu\lambda_1\dots\lambda_s}&=&{1\over s}
\sum_{i=1}^s\left( A_{\lambda_i\mu\lambda_1\dots\lambda_{i-1}\lambda_{i+1}\dots\lambda_s}
\right)
-{s-1\over s}A_{\mu\lambda_1\dots\lambda_s},
\eeqa
illuminates the fact that the duality map serves as a similarity transformation between two representations
of the same algebra \cite{Savvidy-Guttenberg: 2008}. Nevertheless, a bug in the definition of the transposition operator which transforms the fields $A_{\mu\lambda_1\dots\lambda_s}$ into the fields $\tilde{A}_{\mu\lambda_1\dots\lambda_s}$ and in particular the fact that the former is not an idempotent, resulted in the 
necessity for a second version of the dual gauge transformations which would resolve
this bug. 
\subsubsection {2nd Version}
In the article \cite{Savvidy-Guttenberg: 2008} it has been shown that if, instead of (\ref{transpose1}), we use the following definition for the transposed fields,
\beqa\label{transpose1}
\tilde{A}_{\mu\lambda}&=&A_{\lambda\mu}\nn \\
\tilde{A}_{\mu\lambda_1\lambda_2}&=&
{2\over 3}\left(A_{\lambda_1\mu\lambda_2}+A_{\lambda_2\mu\lambda_1}\right)
-{1\over 3}A_{\mu\lambda_1\lambda_2}\nn \\
&\dots\dots&\nn \\
\tilde{A}_{\mu\lambda_1\dots\lambda_s}&=&{2\over s+1}
\sum_{i=1}^s\left( A_{\lambda_i\mu\lambda_1\dots\lambda_{i-1}\lambda_{i+1}\dots\lambda_s}
\right)
-{s-1\over s+1}A_{\mu\lambda_1\dots\lambda_s},
\eeqa
we see that the transposition operator becomes an idempotent, as it should be. 
However, the change in the definition of the transposition operator reflects itself upon the definition of the
gauge transformations, which now acquire the form \cite{Savvidy-Guttenberg: 2008},
\beqa\label{transpose2}
\tilde{\delta}_\xi A_\mu&=&\partial_\mu\xi-ig[A_\mu,\xi]\nn \\
\tilde{\delta}_\xi A_{\mu\lambda}&=&\partial_\lambda\xi_\mu-ig[A_\lambda,\xi_\mu]-ig[A_{\mu\lambda},\xi]\nn \\
\tilde{\delta}_\xi A_{\mu\lambda_1\lambda_2}&=&{2\over 3}\big(
\partial_{\lambda_1}\xi_{\mu\lambda_2}-ig[A_{\lambda_1},\xi_{\mu\lambda_2}]+
\partial_{\lambda_2}\xi_{\mu\lambda_1}-ig[A_{\lambda_2},\xi_{\mu\lambda_1}]\big)
-{1\over 3}\big(
\partial_\mu\xi_{\lambda_1\lambda_2}-ig[A_\mu,\xi_{\lambda_1\lambda_2}]\big)\nn \\
&&-ig{2\over 3}[A_{\mu\lambda_1},\xi_{\lambda_2}]-
ig{2\over 3}[A_{\lambda_1\lambda_2},\xi_\mu]+ig{1\over 3}[A_{\lambda_1\mu},\xi_{\lambda_2}]-\nn \\
&&-ig{2\over 3}[A_{\mu\lambda_2},\xi_{\lambda_1}]-
ig{2\over 3}[A_{\lambda_2\lambda_1},\xi_\mu]+ig{1\over 3}[A_{\lambda_2\mu},\xi_{\lambda_1}]-ig[A_{\mu\lambda_1\lambda_2},\xi]\nn \\
&\dots\dots&
\eeqa
and has been proven to form a closed algebraic structure. We see here that the price we paid for
casting the transposition operator an idempotent, was to include, in the transformations of higher rank fields,
derivatives over the index $\mu$.

\subsection {Symmetrized Extended Gauge Transformations}
According to {\it E. Wigner}, one particle states fall into irreducible representations of the {\it Poincar\'e} group \cite {Wigner: 1939, Bargmann-Wigner: 1948}. Since the irreducible component of a tensor, that describe the physical propagating modes, is its symmetric component, it sounds reasonable to specialise the gauge transformations for 
symmetric, over all their indices, tensor gauge fields \cite{Savvidy: 2006}. 
To simplify the notation, we replace the index $\mu$ with the index $\lambda_1$,
so that the symmetry properties of the tensor become more apparent. The totally symmetric version of 
the transformations (\ref{extended gauge transformations}) is given below, 
\beqa\label{symmetrized extended gauge transformations}
\bar{\delta}_{\xi}A_{\lambda_1\lambda_2}^S&=&\partial_{\lambda_1}\xi_{\lambda_2}+\partial_{\lambda_2}\xi_{\lambda_1}
-ig\left( [A_{\lambda_1}, \xi_{\lambda_2}]+[A_{\lambda_2}, \xi_{\lambda_1}]+[A_{\lambda_1\lambda_2}^S, \xi]\right) \nn \\
\bar{\delta}_{\xi}A_{\lambda_1\lambda_2\lambda_3}^S&=&\partial_{\lambda_1}\xi_{\lambda_2\lambda_3}+
\partial_{\lambda_2}\xi_{\lambda_3\lambda_1}+\partial_{\lambda_3}\xi_{\lambda_1\lambda_2}
-ig\big(
[A_{\lambda_1},\xi_{\lambda_2\lambda_3}]+
[A_{\lambda_2},\xi_{\lambda_3\lambda_1}]+[A_{\lambda_3},\xi_{\lambda_1\lambda_2}]
+\nn \\
&&+[A_{\lambda_2\lambda_3}^S,\xi_{\lambda_1}]
+[A_{\lambda_3\lambda_1}^S,\xi_{\lambda_2}]
+[A_{\lambda_1\lambda_2}^S,\xi_{\lambda_3}]
+[A_{\lambda_1\lambda_2\lambda_3}^S,\xi]
\big)\nn \\
&\dots\dots&\nn \\
\bar{\delta}_{\xi}A_{\lambda_1\dots\lambda_s}^S&=&
\sum_{i=1}^{s}
\partial_{\lambda_i}\xi_{\lambda_1\dots\lambda_{i-1}\lambda_{i+1}\dots\lambda_s}-ig\Big(
\sum_{i=1}^{s}
[A_{\lambda_i},\xi_{\lambda_1\dots\lambda_{i-1}\lambda_{i+1}\dots\lambda_s}]+\nn \\
&&+\sum_{i<j}^s
[A_{\lambda_i\lambda_j}^S,\xi_{\lambda_1\dots\lambda_{i-1}\lambda_{i+1}\dots\lambda_{j-1}
\lambda_{j+1}\dots\lambda_s}]
+\dots+[A_{\lambda_1\dots\lambda_s}^S,\xi]\Big)
\eeqa
The above transformations have been proven to form another set of a closed algebra \cite{Savvidy: 2006}, 
\beqa
[\bar{\delta}_{\xi},\bar{\delta}_{\eta}]A_{\lambda_1\lambda_2\dots\lambda_s}^S=
-ig\bar{\delta}_{\zeta}A_{\lambda_1\lambda_2\dots\lambda_s}^S,
\eeqa
where
\beqa
\zeta&=&[\xi,\eta]\nn \\
\zeta_\lambda&=&[\xi,\eta_\lambda]+[\xi_\lambda,\eta]\nn \\
\zeta_{\lambda_1\lambda_2}&=&[\xi,\eta_{\lambda_1\lambda_2}]
+[\xi_{\lambda_1},\eta_{\lambda_2}]+[\xi_{\lambda_2},\eta_{\lambda_1}]+
[\xi_{\lambda_1\lambda_2}, \eta]\nn \\
&\dots\dots&\nn \\
\zeta_{\lambda_1\dots\lambda_s}&=&[\xi,\eta_{\lambda_1\dots\lambda_s}]+
\sum_{i=1}^s{[\xi_{\lambda_i},\eta_{\lambda_1\dots\lambda_{i-1}\lambda_{i+1}\dots\lambda_s}]}+\dots+
[\xi_{\lambda_1\dots\lambda_s},\eta]
\eeqa
Defining, 
\beqa
A^S_{\mu\lambda_1\dots\lambda_s}\equiv
\sum_{i=1}^sA_{\lambda_i\lambda_1\dots\lambda_{i-1}\mu\lambda_{i+1}\dots\lambda_s},
\eeqa
we observe\footnote{Here we follow the definition of the transpose tensor gauge fields
as given in the 1st version.} that \cite{Savvidy-Barret: 2007},
\beqa
A_{\mu\lambda_1\dots\lambda_s}+\tilde{A}_{\mu\lambda_1\dots\lambda_s}={1\over s}
A^S_{\mu\lambda_1\dots\lambda_s}
\eeqa
Nevertheless, from the above transformations, no homogeneously gauge transformed field strength tensor
can be defined and hence, no gauge invariant {\it Lagrangian} that consists solely of symmetric
tensor gauge fields can be constructed. This is probably a hint that the construction of a higher spin theory dictates for a further departure of the common practices which focus primarily on irreducible representations of the {\it Poincar\'e} group
and hence, on one particle states. This is, however, the case with the {\it Dirac} equation which, inevitably describes
both the particle and antiparticle in the same 4-spinor, reducible representation.
The above observation underlines the central role played by the index $\mu$ in the standard transformations, without which the construction of a gauge invariant {\it Lagrangian} would be impossible. The further relaxation of the 
symmetry of the indices and in particular the formulation of the theory without any symmetry properties between them was 
the primary motivation for the current study.  

\subsection {Conjugate Extended Gauge Transformations}
Here, we review the algebra found in \cite{Savvidy-Georgiou: 2010} and served as the basis for the construction of new topological invariants and {\it Chern-Simons} forms in higher dimensions
\cite{Antoniadis-Savvidy: 2014, Spyros-Savvidy: 2014}. The higher rank tensors fields and tensor gauge functions are antisymmetric under permutations of indices of the same letter, e.g. ($\sigma_1 \leftrightarrow \sigma_2$) and symmetric under permutations of pairs of indices of different letters (e.g. $(\sigma_1\sigma_2)\leftrightarrow (\rho_1\rho_2)$). There is no symmetry with respect to the index $\mu$. 
\beqa\label{conjugate transformations}
\delta A_\mu&=&\partial_\mu\xi-ig[A_\mu,\xi]\nn \\
\delta A_{\mu\sigma_1\sigma_2}&=&\partial_\mu\xi_{\sigma_1\sigma_2}-ig\big([A_\mu, \xi_{\sigma_1\sigma_2}]+
[A_{\mu\sigma_1\sigma_2}, \xi]\big)\nn \\
\delta A_{\mu\sigma_1\sigma_2\rho_1\rho_2}&=&\partial_\mu\xi_{\sigma_1\sigma_2\rho_1\rho_2}
-ig\big([A_\mu, \xi_{\sigma_1\sigma_2\rho_1\rho_2}]+
[A_{\mu\sigma_1\sigma_2}, \xi_{\rho_1\rho_2}]+[A_{\mu\rho_1\rho_2}, \xi_{\sigma_1\sigma_2}]+\nn \\
&&+[A_{\mu\sigma_1\sigma_2\rho_1\rho_2}, \xi]
\big)\nn \\
&\dots\dots&
\eeqa
The above transformations follow the spirit of the standard ones. Indeed, if we treat the additional pairs of indices, as one
index, $\sigma_1\sigma_2=\hat{\sigma}$, the two algebras coincide. 
By construction, only tensor fields of odd number of indices participate in the algebra. 
This, as it might seems restrictive, has been proven sufficient for the construction of new 
topological invariants \cite{Antoniadis-Savvidy: 2014, Spyros-Savvidy: 2014}.

\section {A Systematic Treatment of Extended Gauge Algebras}
In this section we present a systematic method for constructing algebras of gauge transformations
for arbitrary high rank tensor gauge fields. For every tensor field of a given rank, the gauge transformation
will be stated in a generic way via an {\it ansatz} that contains all the possible terms, with arbitrary
coefficients, and the maximum number of tensor gauge functions. 
The requirement for the closure of the algebra will prove to be restrictive, but, nevertheless,
leave a variety of choices. Properly adjusting the values of the initial coefficients and imposing restrictions
on the gauge functions, one can, one the one hand, recover all the, so far, analysed algebras and, on
the other, construct new ones.

\subsection{Vector Gauge Fields}
We begin with the fundamental building block of the {\it Standard Model}, the 
{\it Y-M} vector gauge transformations. Since we know that the group of gauge transformations
is uniquely defined, we will suffice to present a proof for the closure of the algebra, just to 
clarify the details of the formalism that will be adopted throughout the article. 
The infinitesimal, non-abelian gauge transformations are known to be,
\beqa\label{YM gaige transformation}
\delta_{\xi} A_\mu=\partial_\mu\xi-ig[A_\mu, \xi]
\eeqa
In order to check that the algebra of these transformations has a closed
structure, let us calculate the commutator of two successive operations on the
vector field $A_\mu$. During the process, we take into account that the operators
$\delta_\xi$ and $\delta_\eta$, act only on $A_\mu$, since gauge functions are not themselves 
gauge transformed. We have,
\beqa
{i\over g}[\delta_{\xi}, \delta_{\eta}]A_\mu&=&
{i\over g}\big(
\delta_{\xi}\left(\delta_{\eta}A_\mu\right)-
\delta_{\eta}\left(\delta_{\xi}A_\mu\right)\big)=
[\delta_\xi A_\mu,\eta]-[\delta_\eta A_\mu,\xi]\nn \\
&=&[\partial_\mu\xi,\eta]-[\partial_\mu\eta,\xi]-
ig\big([A_\mu,\xi],\eta]-[A_\mu,\eta],\xi]\big)\nn \\
&=&\partial_\mu[\xi,\eta]-ig\big[A_\mu, \left[\xi,\eta\right]\big],
\eeqa
where in the last step we used the {\it Jacobi Identity}. We conclude that, 
\beqa
[\delta_{\xi}, \delta_{\eta}]A_\mu=-ig\delta_\zeta A_\mu,
\eeqa
where $\zeta=[\xi, \eta]$. 

This shows that the commutator of two successive {\it Y-M} gauge transformations 
results in a third one, thus proving that they form a closed algebraic structure. 
\subsection{Second Rank Tensor Gauge Fields}
We wish to follow an analogous procedure to examine the structure of the extended non-abelian 
gauge transformations. In order to begin with the most general transformations for 2nd rank tensor gauge fields,
we introduce two vector gauge functions, $\xi^1_\mu,~\xi^2_\mu$, four coefficients, $c_i$, and begin with the 
{\it ansatz},
\beqa\label{ansatz}
\delta_{\xi} A_{\mu\nu}=\partial_\mu\xi^1_\nu+\partial_\nu\xi^2_\mu-ig\left(
[A_\mu, c_1\xi^1_\nu+ c_2 \xi^2_\nu]+
[A_\nu, c_3\xi^1_\mu+ c_4 \xi^2_\mu]+
[A_{\mu\nu}, \xi]\right).
\eeqa
We shall attempt a determination of the coefficients under the restriction of the
closure of the algebra. The commutator of two successive operations on the
tensor field $A_{\mu\nu}$ is, 
\beqa
&&{i\over g}\left[\delta_{\xi}, \delta_{\eta}\right]A_{\mu\nu}=
{i\over g}\big(\delta_{\xi}(\delta_{\eta}A_{\mu\nu})-\delta_{\eta}(\delta_{\xi}A_{\mu\nu})\big)=
\nn \\
&&=
\left[\delta_\xi A_\mu, c_1\eta^1_\nu+c_2\eta^2_\nu\right]+
\left[\delta_\xi A_\nu, c_3\eta^1_\mu+c_4\eta^2_\mu\right]+
\left[\delta_\xi A_{\mu\nu}, \eta\right]-\nn \\
&&~~~~-
\left[\delta_\eta A_\mu, c_1\xi^1_\nu+c_2\xi^2_\nu\right]-
\left[\delta_\eta A_\nu, c_3\xi^1_\mu+c_4\xi^2_\mu\right]-
\left[\delta_\eta A_{\mu\nu}, \xi\right]
\eeqa 
To determine the coefficients let us first examine the structure of the zeroth order terms over the gauge coupling $g$. 
Substituting (\ref{ansatz}), we get,
\beqa\label{algebra 2nd rank}
&&{i\over g}\left[\delta_{\xi}, \delta_{\eta}\right]A_{\mu\nu}=
[\partial_\mu \xi, c_1\eta^1_\nu+c_2\eta^2_\nu]+
[\partial_\nu \xi, c_3\eta^1_\mu+c_4\eta^2_\mu]+
[\xi, \partial_\mu \eta^1_\nu+\partial_\nu \eta^2_\mu]-\nn \\
&&~~-
[\partial_\mu \eta, c_1\xi^1_\nu+c_2\xi^2_\nu]-
[\partial_\nu \eta, c_3\xi^1_\mu+c_4\xi^2_\mu]-
[\eta, \partial_\mu \xi^1_\nu+\partial_\nu \xi^2_\mu]+\CO(g)=\nn \\
&&=\partial_\mu\left([\xi, \eta^1_\nu]-[\eta, \xi^1_\nu]\right)+
\partial_\nu\left([\xi, \eta^2_\mu]-[\eta, \xi^2_\mu]\right)+\nn \\
&&~~+
[\partial_\mu\xi, c_2\eta^2_\nu+(c_1-1)\eta^1_\nu]-
[\partial_\mu\eta, c_2\xi^2_\nu+(c_1-1)\xi^1_\nu]+\nn \\
&&~~+
[\partial_\nu\xi, c_3\eta^1_\mu+(c_4-1)\eta^2_\mu]-
[\partial_\nu\eta, c_3\xi^1_\mu+(c_4-1)\xi^2_\mu]+\CO(g),
\eeqa
where, 
\beqa
\CO(g)&=&-ig\Big\{
\left[[A_\mu,\xi],c_1\eta^1_\nu+c_2\eta^2_\nu\right]+
\left[[A_\nu,\xi],c_3\eta^1_\mu+c_4\eta^2_\mu\right]-\nn \\
&&~~~~~~-
\left[[A_\mu,\eta],c_1\xi^1_\nu+c_2\xi^2_\nu\right]-
\left[[A_\nu,\eta],c_3\xi^1_\mu+c_4\xi^2_\mu\right]+\nn \\
&&~~~~~~+
\left[[A_\mu,c_1\xi^1_\nu+c_2\xi^2_\nu],\eta\right]+
\left[[A_\nu,c_3\xi^1_\mu+c_4\xi^2_\mu],\eta\right]+
\left[[A_{\mu\nu},\xi],\eta\right]-\nn \\
&&~~~~~~-
\left[[A_\mu,c_1\eta^1_\nu+c_2\eta^2_\nu],\xi\right]-
\left[[A_\nu,c_3\eta^1_\mu+c_4\eta^2_\mu],\xi\right]-
\left[[A_{\mu\nu},\eta],\xi\right]
\Big\}
\eeqa
What is obvious from (\ref{ansatz}) is that if the commutator of two successive transformations is to 
represent another transformation, the zeroth order terms in $g$ must be written solely as
partial derivatives over the two indices of the 2nd rank tensor. Hence, the only choices for the 
terms of the last two lines of (\ref{algebra 2nd rank}) are either to eliminate them by properly choosing
the coefficients $c_i$, or to absorb them in the terms of the first two lines, by imposing suitable
symmetry properties and restricting the vector gauge functions $\xi^i_\mu$.

Following the first option, we keep the two vector gauge functions $\xi^i_\mu,~i=1, 2$, independent 
and eliminate the last four terms of (\ref{algebra 2nd rank}) by setting 
$$c_1=c_4=1~~,~~c_2=c_3=0$$ Then we get,
\beqa
{i\over g}\left[\delta_{\xi}, \delta_{\eta}\right]A_{\mu\nu}=
\partial_\mu\zeta^1_\nu+
\partial_\nu\zeta^2_\mu+\CO(g)
\eeqa
where,
\beqa
\zeta^i_\mu&=&[\xi, \eta^i_\mu]+[\xi^i_\mu, \eta]~~,~~i=1, 2
\eeqa
In order to guarantee that the algebra closes for this specific choice of the $c_is$ and the new 
gauge functions $\zeta^1_\mu,~\zeta^2_\mu$, and furthermore to determine $\zeta$, we need to 
examine the $\CO(g)$ terms. We have,
\beqa
{i\over g}\CO(g)&=&
[[A_\mu, \xi], \eta^1_\nu]+[[\eta^1_\nu, A_\mu], \xi]~+~
[[A_\nu, \xi], \eta^2_\mu]+[[\eta^2_\mu, A_\nu],  \xi]+\nn \\
&&+
[[A_\mu, \xi^1_\nu], \eta]+[[\eta, A_\mu], \xi^1_\nu]~+~
[[A_\nu, \xi^2_\mu], \eta]+[[\eta, A_\nu], \xi^2_\mu]+\nn \\
&&+[[A_{\mu\nu}, \xi], \eta]+
[[\eta, A_{\mu\nu}], \xi]=\nn \\
&=&[A_{\mu}, [\xi, \eta^1_\nu]-[\eta, \xi^1_\nu]]+
[A_{\nu}, [\xi, \eta^2_\mu]-[\eta, \xi^2_\mu]]+
[A_{\mu\nu}, [\xi, \eta]],
\eeqa
where in the last step we employed {\it Jacobi Identity}, once for each pair of adjacent terms. 
We conclude that,
\beqa
\CO(g)=-ig\left\{[A_{\mu},\zeta^1_\nu]+[A_{\nu},\zeta^2_\mu]+
[A_{\mu\nu}, \zeta]\right\},
\eeqa
with $\zeta=[\xi, \eta]$. Hence, the gauge transformations,
\beqa\label{new algebra 2nd rank}
\delta_\xi A_{\mu\nu}=\partial_\mu\xi^1_\nu+\partial_\nu\xi^2_\mu-ig\left(
[A_\mu, \xi^1_\nu]+[A_\nu, \xi^2_\mu]+[A_{\mu\nu}, \xi]
\right),
\eeqa
form a closed group algebra with
\beqa
[\delta_\xi, \delta_\eta]A_{\mu\nu}=
-ig\delta_\zeta A_{\mu\nu}
\eeqa
and
\beqa
\zeta&=&[\xi, \eta]\nn \\
\zeta^1_\mu&=&[\xi, \eta^1_\mu]+[\xi^1_\mu, \eta]\nn \\
\zeta^2_\mu&=&[\xi, \eta^2_\mu]+[\xi^2_\mu, \eta]
\eeqa
What we observe from (\ref{new algebra 2nd rank}) is that the vector field, $A_\mu$, participates in the
gauge transformation of $A_{\mu\nu}$ only as part of the covariant derivative over the two gauge functions\footnote{The standard geometrical interpretation of $A_\mu$, as a principal bundle connection is not directly
extrapolated for analogous interpretations of the higher rank gauge fields. This will be examined in subsequent 
studies.}. 

Thus far, we have managed to merge the two independent, standard and dual, extended gauge transformations
into a more general one, indicating a higher symmetry for the system of 2nd rank tensor gauge fields. 
Before we  continue the same procedure for the 3rd rank tensor gauge fields, let us recover 
the algebras of the standard (\ref{extended gauge transformations}) and  dual (\ref{dual}) transformations 
for the 2nd rank tensor gauge fields.

Let us return to (\ref{algebra 2nd rank}) and check if there is an alternative way to nullify the terms that
cannot be written as partial derivatives with respect to the indices $\mu$ and $\nu$.
If we set the second vector gauge function equal to zero, we see that the algebra closes
if and only if, $c_1=1$, $c_3=0$. This, obviously, recovers the standard transformations 
for the 2nd rank field (\ref{extended gauge transformations}).
If, on the other hand, we set the first vector gauge function to equal to zero, the algebra closes
provided, $c_2=0$, $c_4=1$. This recovers the dual transformations for the 2nd rank field (\ref{dual}).
Lastly, setting in the two vector gauge functions equal, 
and symmetrising the tensor gauge field over its two indices, we recover the first equation of the
symmetrized transformations (\ref{symmetrized extended gauge transformations}).

\subsection{Third Rank Tensor Gauge Fields}
In an analogous way followed in the case of the 2nd rank tensor gauge fields, let us 
introduce three 2nd rank gauge functions $\xi^i_{\mu\nu}, ~i=1,2,3$, and begin with the 
{\it ansatz}\footnote{Directed from the case of the 2nd rank field, the non-homogenous, derivative terms 
combine with the first $\CO(g)$ terms to give the covariant derivative of the higher rank
gauge functions. With this in mind, we do not add coefficients on the first three $\CO(g)$ terms.},
\beqa\label{ansatz3}
&&\delta_\xi A_{\mu\nu\lambda}=\partial_\mu\xi^1_{\nu\lambda}+
\partial_\nu\xi^2_{\lambda\mu}+\partial_\lambda\xi^3_{\mu\nu}-ig\Big(
[A_\mu, \xi^1_{\nu\lambda}]+[A_\nu, \xi^2_{\lambda\mu}]+[A_\lambda, \xi^3_{\mu\nu}]+\nn \\
&&~~~~~~~~~~~+[A_{\mu\nu}, c_1\xi^1_\lambda+c_2\xi^2_\lambda]+
[A_{\nu\lambda}, c_3\xi^1_\mu+c_4\xi^2_\mu]+
[A_{\lambda\mu}, c_5\xi^1_\nu+c_6\xi^2_\nu]+
[A_{\mu\nu\lambda}, \xi]\Big)\nn \\
\eeqa
We will try to determine the 6 coefficients $c_i$ so that the algebra of the gauge transformations forms
a closed structure. We need to underline that the 3rd rank tensor gauge field, $A_{\mu\nu\lambda}$, 
contrary to its standard definition \cite{Savvidy: 2005}, bares no symmetry under the permutation of its last two indices.
The same holds for the three tensor gauge functions, $\xi^i_{\nu\lambda}$ which are no longer symmetric. This is a significant departure which will hopefully lead to an enhancing of the symmetry of the theory
of non-abelian tensor gauge fields. 

The commutator of two successive gauge transformations on the field $A_{\mu\nu\lambda}$ gives,
\beqa
{i\over g}[\delta_\xi,\delta_\eta]A_{\mu\nu\lambda}&=&{i\over g}\big(\delta_{\xi}(\delta_{\eta}A_{\mu\nu\lambda})-\delta_{\eta}(\delta_{\xi}A_{\mu\nu\lambda})\big)=\nn \\
&=&[\delta_{\xi}A_\mu, \eta^1_{\nu\lambda}]+
[\delta_{\xi}A_\nu, \eta^2_{\lambda\mu}]+
[\delta_{\xi}A_\lambda, \eta^3_{\mu\nu}]-
[\delta_{\eta}A_\mu, \xi^1_{\nu\lambda}]-
[\delta_{\eta}A_\nu, \xi^2_{\lambda\mu}]-
[\delta_{\eta}A_\lambda, \xi^3_{\mu\nu}]+
\nn \\
&&+[\delta_\xi A_{\mu\nu}, c_1\eta^1_\lambda+c_2\eta^2_\lambda]+
[\delta_\xi A_{\nu\lambda}, c_3\eta^1_\mu+c_4\eta^2_\mu]+
[\delta_\xi A_{\lambda\mu}, c_5\eta^1_\nu+c_6\eta^2_\nu]-\nn \\
&&-[\delta_\eta A_{\mu\nu}, c_1\xi^1_\lambda+c_2\xi^2_\lambda]-
[\delta_\eta A_{\nu\lambda}, c_3\xi^1_\mu+c_4\xi^2_\mu]-
[\delta_\eta A_{\lambda\mu}, c_5\xi^1_\nu+c_6\xi^2_\nu]+\nn \\
&&+[\delta_\xi A_{\mu\nu\lambda}, \eta]-[\delta_\eta A_{\mu\nu\lambda}, \xi]
\eeqa
To determine the coefficients for the gauge transformation (\ref{ansatz3}), we will first examine the 
structure of the zeroth order terms over the gauge coupling $g$. As is indicated in (\ref{ansatz3}) a 
necessary condition for the closure of the algebra is that the zeroth order terms must be written 
as partial derivatives over the three indices $\mu,~\nu$ and $\lambda$. Focusing on the zeroth order
terms over the coupling $g$, and substituting (\ref{YM gaige transformation}), (\ref{new algebra 2nd rank}) and
(\ref{ansatz3}) in the above equation\footnote{
The substitution of (\ref{new algebra 2nd rank}) does not harm generality since it is the most general transformation
for 2nd rank fields which is compatible with the closure of the algebra.}, one gets,
\beqa
{i\over g}[\delta_\xi,\delta_\eta]A_{\mu\nu\lambda}
&=&\left[\partial_\mu\xi, \eta^1_{\nu\lambda}\right]+
\left[\partial_\nu\xi, \eta^2_{\lambda\mu}\right]+
\left[\partial_\lambda\xi, \eta^3_{\mu\nu}\right]-
\left[\partial_\mu\eta, \xi^1_{\nu\lambda}\right]-
\left[\partial_\nu\eta, \xi^2_{\lambda\mu}\right]-
\left[\partial_\lambda\eta, \xi^3_{\mu\nu}\right]+
\nn \\
&&+
\left[\partial_\mu\xi^1_\nu+\partial_\nu\xi^2_\mu, c_1\eta^1_\lambda+c_2\eta^2_\lambda\right]+
\left[\partial_\nu\xi^1_\lambda+\partial_\lambda\xi^2_\nu, c_3\eta^1_\mu+c_4\eta^2_\mu\right]
+\nn \\
&&+
\left[\partial_\lambda\xi^1_\mu+\partial_\mu\xi^2_\lambda, c_5\eta^1_\nu+c_6\eta^2_\nu\right]
-\left[\partial_\mu\eta^1_\nu+\partial_\nu\eta^2_\mu, c_1\xi^1_\lambda+c_2\xi^2_\lambda\right]-\nn \\
&&-\left[\partial_\nu\eta^1_\lambda+\partial_\lambda\eta^2_\nu, c_3\xi^1_\mu+c_4\xi^2_\mu\right]-
\left[\partial_\lambda\eta^1_\mu+\partial_\mu\eta^2_\lambda, c_5\xi^1_\nu+c_6\xi^2_\nu\right]+\nn \\
&&+
[\partial_\mu\xi^1_{\nu\lambda}+\partial_\nu\xi^2_{\lambda\mu}+\partial_\lambda\xi^3_{\mu\nu}, \eta]-
[\partial_\mu\eta^1_{\nu\lambda}+\partial_\nu\eta^2_{\lambda\mu}+\partial_\lambda\eta^3_{\mu\nu}, \xi]+
\CO(g)=\nn \\
&=&
\partial_\mu\big([\xi,\eta^1_{\nu\lambda}]-[\eta,\xi^1_{\nu\lambda}]\big)
+\partial_\nu\big([\xi,\eta^2_{\lambda\mu}]-[\eta,\xi^2_{\lambda\mu}]\big)+
\partial_\lambda\big([\xi,\eta^3_{\mu\nu}]-[\eta, \xi^3_{\mu\nu}]\big)+\nn \\
&&+c_1\left(
\left[\partial_\mu\xi^1_\nu+\partial_\nu\xi^2_\mu, \eta^1_\lambda\right]-
\left[\partial_\mu\eta^1_\nu+\partial_\nu\eta^2_\mu, \xi^1_\lambda\right]
\right)+\nn\\
&&+c_2\left(
\left[\partial_\mu\xi^1_\nu+\partial_\nu\xi^2_\mu, \eta^2_\lambda\right]-
\left[\partial_\mu\eta^1_\nu+\partial_\nu\eta^2_\mu, \xi^2_\lambda\right]
\right)+\nn \\
&&+c_3\left(
\left[\partial_\nu\xi^1_\lambda+\partial_\lambda\xi^2_\nu, \eta^1_\mu\right]-
\left[\partial_\nu\eta^1_\lambda+\partial_\lambda\eta^2_\nu, \xi^1_\mu\right]
\right)+\nn \\
&&+c_4\left(
\left[\partial_\nu\xi^1_\lambda+\partial_\lambda\xi^2_\nu, \eta^2_\mu\right]-
\left[\partial_\nu\eta^1_\lambda+\partial_\lambda\eta^2_\nu, \xi^2_\mu\right]
\right)+\nn \\
&&+c_5\left(
\left[\partial_\lambda\xi^1_\mu+\partial_\mu\xi^2_\lambda, \eta^1_\nu\right]-
\left[\partial_\lambda\eta^1_\mu+\partial_\mu\eta^2_\lambda, \xi^1_\nu\right]
\right)+\nn \\
&&+c_6\left(
\left[\partial_\lambda\xi^1_\mu+\partial_\mu\xi^2_\lambda, \eta^2_\nu\right]-
\left[\partial_\lambda\eta^1_\mu+\partial_\mu\eta^2_\lambda, \xi^2_\nu\right]
\right)+\CO(g)
\eeqa
Up to zeroth order, over the gauge coupling, we have nine terms in total. The first three have exactly the desired property
for the closure of the algebra. We can isolate the parts of the last six terms, that contain commutators of both the 
vector gauge functions and can be written as partial differentials over 
the three indices, and insert them in the first three terms\footnote{This will simplify the subsequent calculations and especially the cases where one of the vector gauge functions is neglected}. 
\beqa\label{algebra 3rd rank semi final}
{i\over g}[\delta_\xi,\delta_\eta]A_{\mu\nu\lambda}&=&
\partial_\mu\Big([\xi,\eta^1_{\nu\lambda}]-[\eta, \xi^1_{\nu\lambda}]+
c_2\big([\xi^1_\nu,\eta^2_\lambda]-[\eta^1_\nu,\xi^2_\lambda]\big)\Big)+\nn \\
&&+\partial_\nu\Big([\xi,\eta^2_{\lambda\mu}]-[\eta,\xi^2_{\lambda\mu}]+
c_1\big([\xi^1_\lambda, \eta^2_\mu]-[\eta^1_\lambda, \xi^2_\mu]\big)\Big)+\nn \\
&&+
\partial_\lambda\Big([\xi,\eta^3_{\mu\nu}]-[\eta, \xi^3_{\mu\nu}]+
c_3\big([\xi^1_\mu, \eta^2_\nu]-[\eta^1_\mu, \xi^2_\nu]\big)\Big)+\nn \\
&&+
c_1\big([\partial_\mu\xi^1_\nu,\eta^1_\lambda]-[\partial_\mu\eta^1_\nu,\xi^1_\lambda]\big)+
c_4\big([\partial_\lambda\xi^2_\nu, \eta^2_\mu]-[\partial_\lambda\eta^2_\nu, \xi^2_\mu]\big)+\nn \\
&&+
c_2\big([\partial_\nu\xi^2_\mu,\eta^2_\lambda]-[\partial_\nu\eta^2_\mu,\xi^2_\lambda]\big)+
c_5\big([\partial_\lambda\xi^1_\mu, \eta^1_\nu]-[\partial_\lambda\eta^1_\mu, \xi^1_\nu]\big)+\nn \\
&&+
c_3\big([\partial_\nu\xi^1_\lambda, \eta^1_\mu]-[\partial_\nu\eta^1_\lambda, \xi^1_\mu]\big)+
c_6\big([\partial_\mu\xi^2_\lambda, \eta^2_\nu]-[\partial_\mu\eta^2_\lambda, \xi^2_\nu]\big)+\nn \\
&&+(c_4-c_1)\big([\xi^2_\mu,\partial_\nu\eta^1_\lambda]-[\eta^2_\mu,\partial_\nu\xi^1_\lambda]\big)+\nn \\
&&+
(c_5-c_2)\big([\xi^1_\nu,\partial_\mu\eta^2_\lambda]-[\eta^1_\nu,\partial_\mu\xi^2_\lambda]\big)+\nn \\
&&+
(c_6-c_3)\big([\xi^2_\nu,\partial_\lambda\eta^1_\mu]-[\eta^2_\nu,\partial_\lambda\xi^1_\mu]\big)+\CO(g)
\eeqa
Up to zeroth order, over the gauge coupling, we end up with twelve terms. 
Further, one can see that the symmetric parts of the fourth to the ninth terms can also be isolated and written as partial derivatives over the three indices. For example, one can easily show that the symmetric part, over the indices $\nu$ and $\lambda$, of the forth term can be written as a partial derivative over the index $\mu$,
\beqa
[\partial_\mu\xi^1_{(\nu},\eta^1_{\lambda)}]-[\partial_\mu\eta^1_{(\nu},\xi^1_{\lambda)}]=
{1\over 2}\partial_\mu\big(
[\xi^1_\nu,\eta^1_\lambda]+[\xi^1_\lambda,\eta^1_\nu]
\big)
\eeqa
Such is the case for the remaining terms, symmetrized over the proper indices. 
With these in mind we get, 
\beqa\label{algebra last}
{i\over g}[\delta_\xi,\delta_\eta]A_{\mu\nu\lambda}
&=&
\partial_\mu\Big([\xi,\eta^1_{\nu\lambda}]-[\eta, \xi^1_{\nu\lambda}]+
c_2\big([\xi^1_\nu,\eta^2_\lambda]-[\eta^1_\nu,\xi^2_\lambda]\big)+\nn \\
&&~~~~~~~~~~~~~~~~~+
{c_1\over 2}\big([\xi^1_\nu,\eta^1_\lambda]+[\xi^1_\lambda,\eta^1_\nu]\big)+
{c_6\over 2}\big([\xi^2_\lambda, \eta^2_\nu]+[\xi^2_\nu, \eta^2_\lambda]\big)\Big)+\nn \\
&&+\partial_\nu\Big([\xi,\eta^2_{\lambda\mu}]-[\eta,\xi^2_{\lambda\mu}]+
c_1\big([\xi^1_\lambda, \eta^2_\mu]-[\eta^1_\lambda, \xi^2_\mu]\big)+\nn \\
&&~~~~~~~~~~~~~~~~~+
{c_2\over 2}\big([\xi^2_\mu, \eta^2_\lambda]+[\xi^2_\lambda, \eta^2_\mu]\big)
+{c_3\over 2}\big([\xi^1_\lambda, \eta^1_\mu]+[\xi^1_\mu, \eta^1_\lambda]\big)+\nn \\
&&+
\partial_\lambda\Big([\xi,\eta^3_{\mu\nu}]-[\eta, \xi^3_{\mu\nu}]+
c_3\big([\xi^1_\mu, \eta^2_\nu]-[\eta^1_\mu, \xi^2_\nu]\big)+\nn \\
&&~~~~~~~~~~~~~~~~~+
{c_4\over 2}\big([\xi^2_\nu, \eta^2_\mu]+[\xi^2_\mu, \eta^2_\nu]\big)+
{c_5\over 2}\big([\xi^1_\mu, \eta^1_\nu]+[\xi^1_\nu, \eta^1_\mu]\big)\Big)+
\nn \\
&&+c_1\big([\partial_\mu\xi^1_{[\nu},\eta^1_{\lambda]}]-[\partial_\mu\eta^1_{[\nu},\xi^1_{\lambda]}]\big)+
c_4\big([\partial_\lambda\xi^2_{[\nu}, \eta^2_{\mu]}]-[\partial_\lambda\eta^2_{[\nu}, \xi^2_{\mu]}]\big)+\nn \\
&&+
c_2\big([\partial_\nu\xi^2_{[\mu},\eta^2_{\lambda]}]-[\partial_\nu\eta^2_{[\mu},\xi^2_{\lambda]}]\big)+
c_5\big([\partial_\lambda\xi^1_{[\mu}, \eta^1_{\nu]}]-[\partial_\lambda\eta^1_{[\mu}, \xi^1_{\nu]}]\big)+\nn \\
&&+
c_3\big([\partial_\nu\xi^1_{[\lambda}, \eta^1_{\mu]}]-[\partial_\nu\eta^1_{[\lambda}, \xi^1_{\mu]}]\big)+
c_6\big([\partial_\mu\xi^2_{[\lambda}, \eta^2_{\nu]}]-[\partial_\mu\eta^2_{[\lambda}, \xi^2_{\nu]}]\big)+\nn \\
&&+
(c_4-c_1)\big([\xi^2_\mu,\partial_\nu\eta^1_\lambda]-[\eta^2_\mu,\partial_\nu\xi^1_\lambda]\big)+
(c_5-c_2)\big([\xi^1_\nu,\partial_\mu\eta^2_\lambda]-[\eta^1_\nu,\partial_\mu\xi^2_\lambda]\big)+\nn \\
&&+(c_6-c_3)\big([\xi^2_\nu,\partial_\lambda\eta^1_\mu]-[\eta^2_\nu,\partial_\lambda\xi^1_\mu]\big)+\CO(g)
\eeqa

Now we have to nullify the last 9 terms and there are many ways to do this. 
The forth till ninth terms can be omitted, either by nullifying the 
respective coefficients, or by imposing suitable symmetric properties over the indices of the 
3rd rank tensor gauge field or by restricting one vector gauge function (neglecting it, equating it with the
other etc.). The last three terms vanish
either by equating the suitable pairs of $c_is$, or by equating the two vector gauge functions and imposing
proper symmetries on their indices, or by neglecting one of them. Let us examine the
different cases.

If we keep all the tensor gauge functions independent, and do not assume any symmetry properties on the indices of the
tensor gauge fields, then, for the closure of the algebra, it is necessary to set,
\beqa
c_1=c_2=c_3=c_4=c_5=c_6=0
\eeqa
Then we get,
\beqa
{i\over g}[\delta_\xi, \delta_\eta]A_{\mu\nu\lambda}&=&
\partial_\mu\big([\xi,\eta^1_{\nu\lambda}]-[\eta, \xi^1_{\nu\lambda}]\big)+
\partial_\nu\big([\xi,\eta^2_{\lambda\mu}]-[\eta,\xi^2_{\lambda\mu}]\big)+
\partial_\lambda\big([\xi,\eta^3_{\mu\nu}]-[\eta, \xi^3_{\mu\nu}]\big)\nn \\
\eeqa
Let us now examine the closure of the full algebra of the transformations,
\beqa\label{full gauge transformations}
\delta A_{\mu\nu\lambda}=\partial_\mu\xi^1_{\nu\lambda}+
\partial_\nu\xi^2_{\lambda\mu}+\partial_\lambda\xi^3_{\mu\nu}-ig\big(
[A_\mu, \xi^1_{\nu\lambda}]+[A_\nu, \xi^2_{\lambda\mu}]+[A_\lambda, \xi^3_{\mu\nu}]+
[A_{\mu\nu\lambda}, \xi]\big)
\eeqa
taking into consideration all the higher order terms in $g$. We get,
\beqa
{i\over g}[\delta_\xi, \delta_\eta]A_{\mu\nu\lambda}&=&
[\delta_\xi A_\mu, \eta^1_{\nu\lambda}]+
[\delta_\xi A_\nu, \eta^2_{\lambda\mu}]+
[\delta_\xi A_\lambda, \eta^3_{\mu\nu}]+
[\delta_\xi A_{\mu\nu\lambda}, \eta]-\nn \\
&&-
[\delta_\eta A_\mu, \xi^1_{\nu\lambda}]-
[\delta_\eta A_\nu, \xi^2_{\lambda\mu}]-
[\delta_\eta A_\lambda, \xi^3_{\mu\nu}]-
[\delta_\eta A_{\mu\nu\lambda}, \xi]=\nn \\
&=&
[\partial_\mu\xi-ig[A_\mu, \xi], \eta^1_{\nu\lambda}]+
[\partial_\nu\xi-ig[A_\nu, \xi], \eta^2_{\lambda\mu}]+
[\partial_\mu\xi-ig[A_\lambda, \xi], \eta^3_{\mu\nu}]+\nn \\
&&+
\Big[\partial_\mu\xi^1_{\nu\lambda}+
\partial_\nu\xi^2_{\lambda\mu}+\partial_\lambda\xi^3_{\mu\nu}-ig\big(
[A_\mu, \xi^1_{\nu\lambda}]+[A_\nu, \xi^2_{\lambda\mu}]+[A_\lambda, \xi^3_{\mu\nu}]+
[A_{\mu\nu\lambda}, \xi]\big), \eta\Big]-\nn \\
&&~-[\partial_\mu\eta-ig[A_\mu, \eta], \xi^1_{\nu\lambda}]-
[\partial_\nu\eta-ig[A_\nu, \eta], \xi^2_{\lambda\mu}]-
[\partial_\mu\eta-ig[A_\lambda, \eta], \xi^3_{\mu\nu}]-\nn \\
&&-
\Big[\partial_\mu\eta^1_{\nu\lambda}+
\partial_\nu\eta^2_{\lambda\mu}+\partial_\lambda\eta^3_{\mu\nu}-ig\big(
[A_\mu, \eta^1_{\nu\lambda}]+[A_\nu, \eta^2_{\lambda\mu}]+[A_\lambda, \eta^3_{\mu\nu}]+
[A_{\mu\nu\lambda}, \eta]\big), \xi\Big]=\nn \\
&=&
\partial_\mu\big([\xi,\eta^1_{\nu\lambda}]-[\eta, \xi^1_{\nu\lambda}]\big)+
\partial_\nu\big([\xi,\eta^2_{\lambda\mu}]-[\eta,\xi^2_{\lambda\mu}]\big)+
\partial_\lambda\big([\xi,\eta^3_{\mu\nu}]-[\eta, \xi^3_{\mu\nu}]\big)
-\nn \\
&&-ig\big(
\left[A_\mu, [\xi,\eta^1_{\nu\lambda}]-[\eta, \xi^1_{\nu\lambda}]\right]+
\left[A_\nu, [\xi,\eta^2_{\lambda\mu}]-[\eta,\xi^2_{\lambda\mu}]\right]+
\left[A_\lambda, [\xi,\eta^3_{\mu\nu}]-[\eta, \xi^3_{\mu\nu}]\right]+
\nn \\
&&~~~~~~~~+
\left[A_{\mu\nu\lambda}, [\xi, \eta]\right]
\big),
\eeqa
where in the second step we substituted (\ref{YM gaige transformation}), (\ref{full gauge transformations}) and in
last we employed the {\it Jacobi} Identity. We conclude that,
\beqa
[\delta_\xi, \delta_\eta]A_{\mu\nu\lambda}=
-ig\delta_\zeta A_{\mu\nu\lambda},
\eeqa
with,
\beqa
\zeta^i_{\mu\nu}&=&[\xi,\eta^i_{\nu\lambda}]+[\xi^i_{\nu\lambda}, \eta]~~,~~i=1, 2, 3\nn \\
\zeta&=&[\xi, \eta],
\eeqa
which proves that the transformations (\ref{full gauge transformations}) form a closed algebraic structure.
As in the case of the 2nd rank tensor field, the vector field, $A_\mu$, participates in the
gauge transformation of $A_{\mu\nu\lambda}$ only as part of the covariant derivative over the three gauge functions.

Now, let us explore the case where the second vector gauge function 
and the second and third 2nd rank tensor gauge functions are set to zero, 
\beqa
\xi^2_\mu=\xi^2_{\mu\nu}=\xi^3_{\mu\nu}=0
\eeqa
Also, let us symmetrize the 3rd rank tensor gauge field over its last two indices, $\nu$ and $\lambda$.
The last three, together with the fifth, sixth and ninth terms of (\ref{algebra last}), vanish identically because all of them contain the second vector gauge function ($\xi^2_\mu$ or $\eta^2_\mu$). The fourth term vanishes because of the
symmetrization over the indices $\nu$ and $\lambda$. In order to get rid of the seventh and eighth terms,
it is sufficient to set,
\beqa
c_3=c_5
\eeqa
Then, taking the only surviving 2nd rank tensor gauge function $\xi_{\mu\nu}$ symmetric\footnote{
We renamed the indices $\nu$, $\lambda$ to $\lambda_1$, $\lambda_2$ respectively so that 
the their permuting property is visible.}, 
the gauge transformation (\ref{ansatz3}) simplifies to,
\beqa
&&\delta A_{\mu\lambda_1\lambda_2}=\partial_\mu\xi_{\lambda_1\lambda_2}
-ig\Big(
[A_\mu, \xi_{\lambda_1\lambda_2}]+
c_1\big([A_{\mu\lambda_1}, \xi_{\lambda_2}]+[A_{\mu\lambda_2}, \xi_{\lambda_1}]\big)+\nn \\
&&~~~~~~~~~~~~~~+c_3\big(
[A_{\lambda_1\mu},\xi_{\lambda_2}]+[A_{\lambda_2\mu},\xi_{\lambda_2}]+
[A_{\lambda_1\lambda_2},\xi_\mu]+[A_{\lambda_2\lambda_1},\xi_\mu]
\big)+
[A_{\mu\lambda_1\lambda_2}, \xi]\Big)\nn \\
\eeqa
which for $c_1=1$ and $c_3=0$, coincides with the standard gauge transformation (\ref{extended gauge transformations}).

In an analogous way, the 1st version of the dual transformations for the 3rd rank field can be recovered by setting 
$\xi^1_{\mu\nu}=\xi^2_{\mu\nu}=\xi^1_\mu=0$, symmetrising the third tensor gauge function, 
$\xi^3_{\mu\nu}$, and 3rd rank tensor gauge field over its last two indices. 
Then, the closure of the algebra forces, 
\beqa
c_2=c_4,
\eeqa 
so that the gauge transformation becomes, 
\beqa
\delta_{\xi}A_{\mu\lambda_1\lambda_2}&=&
\partial_{\lambda_1}\xi_{\mu\lambda_2}+\partial_{\lambda_2}\xi_{\mu\lambda_1}
-ig\Big( 
[A_{\lambda_1}, \xi_{\mu\lambda_2}]+[A_{\lambda_2}, \xi_{\mu\lambda_1}]+
c_2\big([A_{\mu\lambda_1}, \xi_{\lambda_2}]+[A_{\mu\lambda_2}, \xi_{\lambda_1}]+\nn \\
&&+[A_{\lambda_1\lambda_2},\xi_\mu]+[A_{\lambda_2\lambda_1},\xi_\mu]\big)+
c_6\big([A_{\lambda_1\mu}\xi_{\lambda_2}]+[A_{\lambda_2\mu}\xi_{\lambda_1}]\big)+
[A_{\mu\lambda_1\lambda_2},\xi]\Big),
\eeqa
which for the particular choice $c_2=1$, $c_6=0$, coincides with (\ref{dual}).

It is not hard to see that the 2nd version of the dual transformation  is recovered if we set,
\beqa
&&\xi^3_{\mu\nu}=0~~,~~\xi^{2}_{\mu\nu}=-4\xi^{1}_{\mu\nu}~~,~~\xi^1_\mu=0~~,~~c_2=c_4=1~~,~~c_6=-1/2,
\eeqa
and symmetrise and the 3rd rank tensor gauge field over its last two indices together with
the surviving 2nd rank tensor gauge functions. 
Then the exact form of (\ref{transpose2}) is recovered if we normalise 
\beqa
\xi^1_{\mu\nu}=-{1\over 3}\xi_{\mu\nu}~~,~~\xi^2_\mu={4\over 3}\xi_\mu
\eeqa

Finally, let us examine the case where the 3rd rank tensor is antisymmetric over
its last two indices. If we also set, 
\beqa
\xi^2_\mu=\xi^2_{\mu\nu}=\xi^3_{\mu\nu}=0
\eeqa
it is easy to see from (\ref{algebra 3rd rank semi final}) that the nullification of the terms that cannot
be written as partial differentials is achieved only if we set, 
\beqa
c_1=0~~~,~~~c_5=-c_3
\eeqa
Then (\ref{ansatz3}) reduces to,
\beqa
\delta_\xi A_{\mu\sigma_1\sigma_2}&=&\partial_\mu\xi_{\sigma_1\sigma_2}-ig\Big(
[A_\mu,\xi_{\sigma_1\sigma_2}]+{c_3\over 2}\big(
[A_{\sigma_1\sigma_2},\xi_\mu]-[A_{\sigma_2\sigma_1},\xi_\mu]+[A_{\sigma_1\mu},\xi_{\sigma_2}]-[A_{\sigma_2\mu},\xi_{\sigma_1}]
\big)+\nn \\
&&+[A_{\mu\sigma_1\sigma_2}, \xi]\Big),
\eeqa
where the tensor gauge function, $\xi_{\mu\nu}$, is antisymmetric under the permutation of its two indices. 
Now the commutator of two successive transformations (\ref{algebra 3rd rank semi final}) reduces to,
\beqa
{i\over g}[\delta_\xi, \delta_\eta]A_{\mu\sigma_1\sigma_2}&=&\partial_\mu\big(
[\xi, \eta_{\sigma_1\sigma_2}]+[ \xi_{\sigma_1\sigma_2}, \eta]\big)+\nn \\
&&+c_3\Big(
\partial_{\sigma_1}\big([\xi_{\sigma_2}, \eta_\mu]+[\xi_\mu, \eta_{\sigma_2}]\big)-
\partial_{\sigma_2}\big([\xi_{\sigma_1}, \eta_\mu]+[\xi_\mu, \eta_{\sigma_1}]\big)\Big)+\CO(g)\nn \\
\eeqa 
Since the closure of the algebra requires that partial derivatives in terms of the indices 
$\sigma_1$ and $\sigma_2$ should be absent when we ignore the gauge functions $\xi^2_{\mu\nu}, \xi^3_{\mu\nu}$, the only legitimate possibility is to set  $c_3=0$. Thus, we recover the algebra 
(\ref{conjugate transformations}) which has been proven to form a closed 
structure \cite{Savvidy-Georgiou: 2010}. 

\subsection {The General Case}
The method we developed seems to be sufficiently generic to accomplish two aims. 
On the one hand to recover all the existing closed algebras of higher rank tensor gauge fields and thus
to embed them in a more general framework, on the other to provide a tool for the investigation of new ones. 
Implementing it as a tool, the existence of a brand new closed algebra for higher rank tensor gauge fields became apparent. Hence, we have proved that up to the tensor of the third tank, the following transformations
provide a closed algebraic structure,
\beqa
\delta_\xi A_\mu&=&\partial_\mu\xi-ig[A_\mu, \xi]\nn \\
\delta_\xi A_{\mu\nu}&=&\partial_\mu\xi^1_\nu+\partial_\nu\xi^2_\mu-ig\left(
[A_\mu, \xi^1_\nu]+[A_\nu, \xi^2_\mu]+[A_{\mu\nu}, \xi]
\right)\nn \\
\delta_\xi A_{\mu\nu\lambda}&=&\partial_\mu\xi^1_{\nu\lambda}+
\partial_\nu\xi^2_{\lambda\mu}+\partial_\lambda\xi^3_{\mu\nu}-ig\Big(
[A_\mu, \xi^1_{\nu\lambda}]+[A_\nu, \xi^2_{\lambda\mu}]+[A_\lambda, \xi^3_{\mu\nu}]+
[A_{\mu\nu\lambda}, \xi]\Big)\nn \\
\eeqa

It seems natural to postulate that the gauge transformations for the general case of a r-th rank gauge field is  
given by,
\beqa\label{general gauge transformations}
\delta_\xi A_{\mu_1\dots\mu_r}&=&
\sum_{i=1}^r\partial_{\mu_i}\xi^i_{\mu_{i+1}\dots\mu_r\mu_1\dots\mu_{i-1}}-ig\Big(
\sum_{i=1}^r[A_{\mu_i}, \xi^i_{\mu_{i+1}\dots\mu_r\mu_1\dots\mu_{i-1}}]+
[A_{\mu_1\dots\mu_r}, \xi]
\Big).\nn \\
\eeqa
Let us prove that the postulated generalisation forms a closed algebra.
The commutator of two successive transformations gives,
\beqa
{i\over g}[\delta_\xi,\delta_\eta]A_{\mu_1\dots\mu_r}&=&{i\over g}\big(\delta_{\xi}(\delta_{\eta}A_{\mu_1\dots\mu_r})-\delta_{\eta}(\delta_{\xi}A_{\mu_1\dots\mu_r})\big)=\nn \\
&=&\sum_{i=1}^r [\delta_\xi A_{\mu_i},\eta^i_{\mu_{i+1}\dots\mu_r\mu_1\dots\mu_{i-1}}]-
\sum_{i=1}^r [\delta_\eta A_{\mu_i},\xi^i_{\mu_{i+1}\dots\mu_r\mu_1\dots\mu_{i-1}}]+\nn \\
&&+[\delta_\xi A_{\mu_1\dots\mu_r},\eta]-[\delta_\eta A_{\mu_1\dots\mu_r},\xi]=\nn \\
&=&\sum_{i=1}^r \left[\partial_{\mu_i}\xi-ig[A_{\mu_i},\xi],\eta^i_{\mu_{i+1}\dots\mu_r\mu_1\dots\mu_{i-1}}\right]-
\sum_{i=1}^r \left[\partial_{\mu_i}\eta-ig[A_{\mu_i},\eta],\xi^i_{\mu_{i+1}\dots\mu_r\mu_1\dots\mu_{i-1}}\right]+\nn \\
&&+
\sum_{i=1}^r\left[
\partial_{\mu_i}\xi^i_{\mu_{i+1}\dots\mu_r\mu_1\dots\mu_{i-1}}-ig\left[A_{\mu_i},\xi^i_{\mu_{i+1}\dots\mu_r\mu_1\dots\mu_{i-1}}\right],\eta\right]
-ig\big[\left[A_{\mu_1\dots\mu_r},\xi\right],\eta\big]-\nn \\
&&-
\sum_{i=1}^r\left[
\partial_{\mu_i}\eta^i_{\mu_{i+1}\dots\mu_r\mu_1\dots\mu_{i-1}}-ig\left[A_{\mu_i},\eta^i_{\mu_{i+1}\dots\mu_r\mu_1\dots\mu_{i-1}}\right],\xi\right]
+ig\big[\left[A_{\mu_1\dots\mu_r},\eta\right],\xi\big]=\nn \\
&=&\sum_{i=1}^r\partial_{\mu_i}\left(
\left[\xi,\eta^i_{\mu_{i+1}\dots\mu_r\mu_1\dots\mu_{r-1}}\right]+
\left[\xi^i_{\mu_{i+1}\dots\mu_r\mu_1\dots\mu_{r-1}},\eta\right]\right)-\nn \\
&&-ig\left\{
\sum_{i=1}^r\left[
A_{\mu_i},\big[\xi,\eta^i_{\mu_{i+1}\dots\mu_i\mu_1\dots\mu_{r-1}}\big]+
\big[\xi^i_{\mu_{i+1}\dots\mu_i\mu_1\dots\mu_{r-1}},\eta\big]
\right]+\big[A_{\mu_1\dots\mu_r},[\xi,\eta]\big]\right\}\nn \\
\eeqa
In the third step we substituted (\ref{YM gaige transformation}) and (\ref{general gauge transformations}) and 
in the last, we employed the {\it Jacobi Identity}, where needed. We conclude that, 
\beqa
[\delta_\xi,\delta_\eta]A_{\mu_1\dots\mu_r}=-ig\delta_\zeta A_{\mu_1\dots\mu_r},
\eeqa
with
\beqa
\zeta&=&[\xi, \eta]\nn \\
\zeta^i_{\mu_{i+1}\dots\mu_r\mu_1\dots\mu_{i-1}}&=&
\left[\xi, \eta^i_{\mu_{i+1}\dots\mu_r\mu_1\dots\mu_{i-1}}\right]+
\left[\xi^i_{\mu_{i+1}\dots\mu_r\mu_1\dots\mu_{i-1}},\eta \right],
\eeqa
hence that the general gauge transformation (\ref{general gauge transformations}) forms a closed algebraic
structure. 

\section{Conclusions}
We presented a general method for constructing extended gauge transformations that include bosonic fields
of arbitrary high spin under the requirement that they should form a closed algebraic structure. 
After we having recovered the known closed algebras of extended gauge transformations (\ref{extended gauge transformations}), (\ref{dual}), (\ref{transpose2}), (\ref{symmetrized extended gauge transformations}) and (\ref{conjugate transformations}),
by properly adjusting the initial coefficients and restricting the tensor gauge functions of the transformations, 
we advocated for the existence of the following new algebra of extended gauge transformations,
\beqa
\delta_\xi A_\mu&=&\partial_\mu\xi-ig[A_\mu, \xi]\nn \\
\delta_\xi A_{\mu\nu}&=&\partial_\mu\xi^1_\nu+\partial_\nu\xi^2_\mu-ig\left(
[A_\mu, \xi^1_\nu]+[A_\nu, \xi^2_\mu]+[A_{\mu\nu}, \xi]
\right)\nn \\
\delta_\xi A_{\mu\nu\lambda}&=&\partial_\mu\xi^1_{\nu\lambda}+
\partial_\nu\xi^2_{\lambda\mu}+\partial_\lambda\xi^3_{\mu\nu}-ig\Big(
[A_\mu, \xi^1_{\nu\lambda}]+[A_\nu, \xi^2_{\lambda\mu}]+[A_\lambda, \xi^3_{\mu\nu}]+
[A_{\mu\nu\lambda}, \xi]\Big)\nn \\
\dots\dots\dots\nn \\
\delta_\xi A_{\mu_1\dots\mu_r}&=&
\sum_{i=1}^r\partial_{\mu_i}\xi^i_{\mu_{i+1}\dots\mu_r\mu_1\dots\mu_{i-1}}-ig\Big(
\sum_{i=1}^r[A_{\mu_i}, \xi^i_{\mu_{i+1}\dots\mu_r\mu_1\dots\mu_{i-1}}]+
[A_{\mu_1\dots\mu_r}, \xi]
\Big),\nn \\
\eeqa
which was proven to be closed under the commutator,
\beqa
[\delta_\xi,\delta_\eta]A_{\mu_1\dots\mu_r}=-ig\delta_\zeta A_{\mu_1\dots\mu_r},
\eeqa
with,
\beqa
\zeta&=&[\xi, \eta]\nn \\
\zeta^i_{\mu_{i+1}\dots\mu_r\mu_1\dots\mu_{i-1}}&=&
\left[\xi, \eta^i_{\mu_{i+1}\dots\mu_r\mu_1\dots\mu_{i-1}}\right]+
\left[\xi^i_{\mu_{i+1}\dots\mu_r\mu_1\dots\mu_{i-1}},\eta \right]
\eeqa
The potential usage of the new algebra is still under investigation. The fact that it introduces the same number of
gauge functions as the rank of the tensor gauge field indicates a higher symmetry for the system
and sounds promising as regards the cancellation of non-propagating degrees of freedom. 
It is worth to mention that at each step of the transformations, apart from the field transformed, no lower rank tensor
fields participate other than the {\it Y-M} field. The latter participate exactly in the way to define the covariant derivative
on each of the tensor gauge functions,
$$
\nabla_{\mu_i}\xi^i_{\mu_{i+1}\dots\mu_r\mu_1\dots\mu_{i-1}}\equiv \partial_{\mu_i}\xi^i_{\mu_{i+1}\dots\mu_r\mu_1\dots\mu_{i-1}}-ig
[A_{\mu_i}, \xi^i_{\mu_{i+1}\dots\mu_r\mu_1\dots\mu_{i-1}}],
$$
hence playing its custom, geometrical role as a principal bundle connection.

\section{Acknowledgements}
The author would like to thank prof. G. Savvidy for his useful remarks and creative discussions
we had together. This article is dedicated to my students in the University of Patras, during the years
2016-2019. Their direct expression of support was invaluable.


\newpage
\begin{thebibliography}{99}

\bibitem{Dirac: 1936}
P. Dirac,
\emph{Relativistic Wave Equations}, 
Proc. Roy. Soc. A 155 (1936), p. 447

\bibitem{Fierz-Pauli: 1939}
M. Fierz, W. Pauli,
\emph{On Relativistic Wave Equations for Particles of Arbitrary Spin in an Electromagnetic Field},
Proc. R. Soc. London A 173 (1939), p. 211.

\bibitem {Wigner: 1939}
E. Wigner, 
\emph{Unitary representations of the inhomogeneous Lorentz group}, 
Annals of Mathematics 40 (1939), pp. 149-204 

\bibitem {De Wet: 1940}
J. De Wet,
\emph{On the Spinor Equations for Particles with Arbitrary Spin and Rest Mass Zero},
Phys. Rev. 58 (1940), p. 236  

\bibitem{Bargmann-Wigner: 1948}
V. Bargmann, E. Wigner,
\emph{Group Theoretical Discussion Of Relativistic Wave Equations},
Proc. Natl. Acad. Sci. USA 34 (1948), pp. 211-223

\bibitem{Yang-Mills: 1954}
C. Yang, R. Mills,
\emph{Conservation of Isotopic Spin and Isotopic Gauge Invariance}, 
Phys. Rev. 96 (1954), pp. 191-195


\bibitem{Fronsdal: 1957}
C. Fronsdal,
\emph{On The Theory of Higher Spin Fields},
Il Nuovo Cimento, vol. 9, issue S2 (1957), pp. 416-443

\bibitem{Fronsdal: 1978}
C. Fronsdal,
\emph{Massless Fields with Integer Spin},
Phys.Rev. D18 (1978), p. 3624

\bibitem{Fang-Fronsdal: 1978}
J. Fang, C. Frodsdal
\emph{Massless Fields with Half Integral Spin},
Phys.Rev. D18 (1978), p. 3630

\bibitem{Savvidy: 2005}
G. Savvidy,
\emph{Non-Abelian Tensor Gauge Fields: Generalization of Yang-Mills theory},
Physics Letters B 625 (2005), pp. 341-350

\bibitem{Savvidy: 2006}
G. Savvidy,
\emph{Non-Abelian Tensor Gauge Fields I},
Int. J. Mod. Phys. A 21 (2006), p. 4931

\bibitem{Savvidy: 2006b}
G. Savvidy,
\emph{Non-Abelian Tensor Gauge Fields II},
Int. J. Mod. Phys. A 21, (2006), pp. 4959-4977

\bibitem {Savvidy: 2006c}
G. Savvidy,
\emph{Non-Abelian Tensor Gauge Fields and Higher-Spin Extension of Standard Model},
Fortschr. Phys. 54 (2006), pp. 472-486

\bibitem{Savvidy-Barret: 2007}
J. Barrett, G. Savvidy,
\emph{A dual Lagrangian for non-Abelian tensor gauge fields},
Physics Letters B 652 (2007), pp. 141-145

\bibitem {Spyros-Savvidy: 2008}
S. Konitopoulos, G. Savvidy,
\emph {Propagating modes of a non-Abelian tensor gauge field of second rank},
J. Phys. A: Math. Theor. 41 (2008), p. 355402


\bibitem{Savvidy-Guttenberg: 2008}
S. Guttenberg, G. Savvidy,
\emph{Duality Transformation of non-Abelian Tensor Gauge Fields},
Modern Physics Letters A 23, 14 (2008), pp. 999-1009

\bibitem{Spyros-Fazio-Savvidy: 2009}
S. Konitopoulos, R. Fazio, G. Savvidy,
\emph{Tensor gauge boson production in high-energy collisions},
EPL, 85 (2009), p. 51001

\bibitem {Spyros-Savvidy: 2009}
S. Konitopoulos, G. Savvidy,
\emph{Production of Charged Spin-Two Gauge Bosons in Gluon-Gluon Scattering},
arXiv: 0812.4345 [hep-th]

\bibitem{Savvidy: 2010}
G. Savvidy,
\emph{Topological mass generation four-dimensional gauge theory},
Physics Letters B 694 (2010), pp. 65-73

\bibitem{Savvidy-Georgiou: 2010}
G. Georgiou,~G. Savvidy,
\emph{Non-abelian Tensor Gauge Fields and New Topological Invariants},
arXiv:1212.5228 (2012) [hep-th]

\bibitem{Savvidy: 2010b}
G. Savvidy,
\emph{Extension of the Poincar\'e Group and non-Abelian Tensor Gauge Fields},
Int. J. Mod. Phys. A 25 (2010), pp. 5765-5785

\bibitem{Antoniadis-Savvidy: 2014}
I. Antoniadis, G. Savvidy,
\emph{Extension of Chern-Simons Forms and New Gauge Anomalies},
Int. J. Mod. Phys. A 29 (2014), p. 1450027

\bibitem{Spyros-Savvidy: 2014}
S. Konitopoulos, G. Savvidy,
\emph{Extension of Chern-Simons Forms},
J. Mat. Phys. 55, (2014), p. 062304

\bibitem{Savvidy: 2014}
G. Savvidy,
\emph{Asymptotic Freedom of Non-Abelian Tensor Gauge Fields},
Physics Letters B 732 (2014), pp. 150-155

\bibitem{Spyros-Savvidy: 2016}
S. Konitopoulos, G. Savvidy,
\emph{Proton structure, its spin and tensor gluons},
EPJ Web of Conferences 125 (2016), p. 04016 


\bibitem{Salgados: 2017}
F. Izaurieta, P. Salgado, S. Salgado,
\emph{Chern-Simons-Antoniadis-Savvidy forms and standard supergravity},
Phys. Lett. B 767 (2017), pp. 360-365



\end{thebibliography}
\end{document}